\documentstyle[epsf]{article}
\newcommand{\pspicture}[1]{
\centerline{\setlength\epsfxsize{14cm}\epsfbox{#1}}
}
\newcommand{\BtoKstargamma}{B \to K^*\gamma}
\newcommand{\btosgamma}{b \to s\gamma}
\newcommand{\GeV}{\mbox{GeV}}
\newcommand{\ukqcdresult}{0.15^{+12}_{-14}}
\newcommand{\ukqcdflatresult}{0.15^{+5}_{-4}}
\newcommand{\cleoresult}{0.21(5)}
\newcommand{\cleoone}{0.23(6)}
\newcommand{\cleotwo}{0.19(5)}
\newcommand{\titletext}{$B \to K^* \gamma$~: Penguins on the Lattice}
\newcommand{\abstracttext}{
We calculate the leading--order matrix element for the decay
$\BtoKstargamma$ in the quenched approximation of lattice QCD on a
$24^3\times48$ lattice at $\beta=6.2$, using an $O(a)$-improved
fermion action.  Extrapolating the quark masses to their physical
values we obtain an on-shell form factor of $T_1(q^2{=}0)=\ukqcdresult$,
where the errors quoted are purely statistical.  
We find $T_1$ is
approximately independent of the spectator quark mass and extract
$T_1(q^2{=}0)=\ukqcdflatresult$ if this independence is assumed.  We
compare this with the same form factor derived (in the Standard Model)
from the CLEO experimental branching ratio of $BR(\BtoKstargamma) =
(4.5 \pm 1.5 \pm 0.9) \times 10^{-5}$ and find the results to be
consistent within statistical errors.}

\title{\titletext}
\author{UKQCD Collaboration}
\begin{document}

\begin{titlepage}

\begin{flushright}
Edinburgh Preprint  93/528\\
Southampton Preprint SHEP 93/94--01 \\
hep-lat version 
\end{flushright}

\vspace*{5mm}

\begin{center}
{\Huge \titletext }\\[15mm]
{\large\it UKQCD Collaboration
}\footnote{{\tt info-ukqcd@hep.ph.soton.ac.uk}}
\\[3mm]

{\bf K.C.~Bowler, D.S.~Henty, N.M.~Hazel, H.~Hoeber,
R.D.~Kenway, D.G.~Richards, H.P.~Shanahan, J.N.~Simone, }\\
Department of Physics, The University of Edinburgh, Edinburgh
EH9~3JZ, Scotland

{\bf J.M.~Flynn, B.J.~Gough, }\\ Physics Department, The
University, Southampton SO9~5NH, UK
\end{center}
\vspace{5mm}
\begin{abstract}
\abstracttext
\end{abstract}

\begin{center}
To be submitted to {\em Phys. Rev. Lett.}
\end{center}

\vfill

\begin{flushleft}
PACS: 12.38.Gc, 13.40.Hq, 14.40.Jz
\end{flushleft}

\end{titlepage}

\maketitle
\begin{abstract}
\abstracttext
\end{abstract}

\section{Introduction}
Theoretical interest in the rare decay $\BtoKstargamma$ as a
test of the Standard Model has recently been renewed by the
experimental results of the CLEO collaboration
\cite{cleo:evidence-for-penguins}. For the first time, this mode
has been positively identified and a preliminary determination
of its branching ratio given.

The significance of $\BtoKstargamma$ arises from the underlying
flavor-changing quark-level process $\btosgamma$, which first
occurs through penguin-type diagrams at one-loop in the Standard
Model.  The process is also sensitive to new physics appearing
as virtual particles in the internal loops. Existing bounds on
the $\btosgamma$ branching ratio have been used to place
constraints on supersymmetry
\cite{diaz:supersymmetry,barger:supersymmetry,borzumati:supersymmetry}
and other extensions of the Standard Model
\cite{rizzo:two-higgs-doublet,hou:fourth-generation}.

In order to compare the experimental branching ratio with a
theoretical prediction it is necessary to know the relevant
hadronic matrix elements. These have been estimated using a wide
range of methods, including relativistic and nonrelativistic
quark models 
\cite{deshpande:rel-quark-model,odonnel:rel-quark-model,altomari:nonrel-quark-model}, 
two-point and three-point QCD sum rules
\cite{dominguez:2pt-sum-rules,aliev:2pt-sum-rules,ball:3pt-sum-rules,colangelo:3pt-sum-rules}
and heavy quark symmetry
\cite{ali:heavy-quark-symmetry}, 
but there remains some disagreement between the different
results. It is therefore of interest to perform a direct
calculation of the matrix elements using lattice QCD. The
viability of the lattice approach was first demonstrated by the
work of Bernard, Hsieh and Soni
\cite{bernard:lattice-91} in 1991.

In the leading-log approximation the $\BtoKstargamma$ transition
is caused by a single chiral magnetic moment operator from the
effective weak Hamiltonian. In the notation of Grinstein,
Springer and Wise
\cite{grinstein:b-meson-decay} this is, 
\begin{equation} 
O_7=\frac{e}{16 \pi^2} m_b \overline{s} \sigma_{\mu\nu}
\frac{1}{2}(1+\gamma_5)b ~F^{\mu\nu}, 
\end{equation}
with an on-shell matrix element given by, 
\begin{equation}
{\cal M}= \frac{e G_F m_b}{2 \sqrt{2} \pi^2} 
C_7(m_b) V_{tb} V_{ts}^*  \eta^{\mu*} \langle K^* |
\overline{s} \sigma_{\mu\nu} q^\nu b_R | B \rangle , 
\end{equation}
where $q$ and $\eta$ are the momentum and polarization of the
emitted photon.  The coefficient $C_7(m_b)$ arises from the
mixing of $O_7$ with other effective operators in running the
scale down from $M_W$ to $m_b$. The anomalous dimension matrix
of all the effective operators at the one-loop level has been
calculated by several groups and is now well-understood
\cite{cho:weak-hamiltonian}.

Following Bernard {\em et al.} the general matrix element can be
parametrized in terms of the momentum, $k$, and polarization,
$\epsilon$, of the $K^*$, and the momentum, $p$, of the $B$ meson,
using three form factors, $T_1$, $T_2$ and $T_3$, where $T_1$ is
chosen to be real, so that $T_2$ and $T_3$ are purely imaginary,
\begin{equation}
 \langle K^*(k,\epsilon) | J_\mu | B(p) \rangle 
=
C^{1}_{\mu}T_1(q^2)
+C^{2}_{\mu} T_2(q^2)
+C^{3}_{\mu} T_3(q^2)
\end{equation}
\begin{equation}
J_\mu = \overline{s} \sigma_{\mu\nu} q^\nu b_R , 
\quad q=p-k,
\end{equation}
\begin{eqnarray}
C^{1}_\mu & = & 
2\varepsilon_{\mu\nu\lambda\rho} \epsilon^\nu p^\lambda k^\rho \\
C^{2}_\mu & = & 
\epsilon_\mu(m_B^2 - m_{K^*}^2) - \epsilon\cdot q (p+k)_\mu \\
C^{3}_\mu & = & 
\epsilon\cdot q 
\left( q_\mu - \frac{q^2}{m_B^2-m_{K^*}^2} (p+k)_\mu \right).
\end{eqnarray}
The on-shell $(q^2=0)$ matrix element depends on $T_1$ only,
since $T_2(q^2{=}0)=-i T_1(q^2{=}0)$ and the coefficient of $T_3$
is zero. Performing the necessary phase space integral and sums
over polarization vectors gives the decay width for
$\BtoKstargamma$,
\begin{equation}
\label{eq:decay-rate}
\Gamma(\BtoKstargamma)= \frac{\alpha}{8 \pi^4} m_b^2 G_F^2
            m_B^3 \left(1-\frac{m_{K^\ast}^2}{m_B^2}\right)^3 
            | V_{tb} V_{ts}^\ast |^2 |C_7(m_b)|^2 |T_1(q^2{=}0)|^2.
\end{equation}
By computing the matrix elements on the lattice for various
$q^2$, the on-shell value of the form factor $T_1(0)$
can be obtained by interpolation.

\section{Computational Details}

We work in the quenched approximation on a $24^3\times 48$
lattice at $\beta=6.2$, which corresponds to an inverse lattice
spacing $a^{-1} = 2.73(5)\,\GeV$, evaluated by measuring the 
string tension
\cite{ukqcd:light-hadron-spectrum}. 
Our calculation is performed on sixty
$SU(3)$ gauge field configurations (for details see
Refs.~\cite{ukqcd:light-hadron-spectrum} and ~\cite{ukqcd:strange}).  
The quark propagators are calculated using an $O(a)$-improved Wilson
fermion action \cite{sw:improved-action}. 
We use
gauge--invariant smeared sources for the heavy quark propagators 
with an r.m.s. smearing radius of $5.2$ \cite{ukqcd:smearing}.
Local sources are used for the light quark propagators.

As the mass of the $b$ quark is almost twice the inverse lattice
spacing, direct computation of a $b$-quark propagator is not feasible.
We therefore compute heavy-quark propagators with masses in the region
of the charm-quark mass, and extrapolate.

Our statistical errors are calculated according to the bootstrap
procedure described in Ref.~\cite{ukqcd:light-hadron-spectrum}, using
250 bootstrap samples.

To obtain the matrix element $\langle V(k) | \overline{s}
\sigma_{\mu\nu} b | P(p) \rangle$, we calculate a ratio of
three-point and two-point correlators,
\begin{equation}  
\label{eq:3pt-correlator}
C_{\rho\mu\nu}(t,t_f,\vec{p},\vec{q}) =
\frac{
C^{3pt}_{\rho\mu\nu}(t,t_f,\vec{p},\vec{q})
}{
C^{2pt}_{P}(t_f-t,\vec{p}) C^{2pt}_{V}(t,\vec{p}-\vec{q})
},
\end{equation}
where,
\[
C^{3pt}_{\rho\mu\nu}(t,t_f,\vec{p},\vec{q})  = 
\sum_{\vec{x},\vec{y}}  
e^{i \vec{p}\cdot \vec{x}}  
e^{- i \vec{q}\cdot \vec{y}} 
\langle J^\dagger_{P}(t_f,\vec{x}) T_{\mu\nu}(t,\vec{y}) 
J_{V\rho}(0) \rangle,
\]
\begin{equation}
 \mathop{\longrightarrow}\limits_{t,t_f - t\to\infty} \sum_{\epsilon} 
\frac{Z_{P}}{2 E_{P}} 
\frac{Z_{V}}{2 E_{V}} 
e^{-E_{P} (t_f-t) } e^{-E_V t} 
\epsilon_\rho 
\langle P(p) | \overline{b} \sigma_{\mu\nu} s | 
 V(k,\epsilon) \rangle,
\end{equation}
and
\begin{eqnarray}
C^{2pt}_{P}(t,\vec{p}) = &
\sum_{\vec{x}} e^{i\vec{p}\cdot\vec{x}}
  \langle J^\dagger_{P}(t,\vec{x}) J^{\vphantom{\dagger}}_{P}(0)
\rangle & \mathop{\rightarrow}\limits_{t\to\infty}
\frac{Z^2_{P}}{2 E_{P}} e^{-E_{P}t} \nonumber \\
C^{2pt}_{V}(t,\vec{k}) = & -{\displaystyle 
\left(\frac{1}{3}\right)}
 \sum_{\vec{x}}  e^{i\vec{k}\cdot\vec{x}}
  \langle J_{V\sigma}^\dagger(t,\vec{x}) 
  J^{\sigma}_{V}(0) \rangle & \mathop{\rightarrow}\limits_{t\to\infty}
\frac{Z^2_{V}}{2 E_V} e^{-E_V t} 
\end{eqnarray}
with $J_{P}$ and $J_V$ interpolating fields for the pseudoscalar and
vector mesons respectively.  $T_{\mu\nu}$ is the $O(a)$-improved
version of the operator $\bar b\sigma_{\mu\nu} s$
\cite{heatlie:clover-action}. The full matrix elements can then be 
derived by using the relation $\sigma_{\mu\nu}\gamma_5 = -\frac{i}{2}
\epsilon_{\mu\nu\lambda\rho} \sigma^{\lambda\rho}$. We employ time
reversal symmetry to obtain the correctly ordered matrix element,
$\langle V(k) | \overline{s} \sigma_{\mu\nu} b | P(p) \rangle$. To
evaluate these correlators, we use the standard source method
\cite{bernard:source-method}. We choose $t_f=24$ and symmetrize the
correlators about that point using Euclidean time reversal
\cite{bernard:folding}.  We evaluate $C_{\rho\mu\nu}$ for three values
of the light quark mass ($\kappa_l=0.14144, 0.14226, 0.14262$),
two values of the strange quark mass ($\kappa_s=0.14144,
0.14226$) which straddle the physical value (given by
$\kappa^{phys}_s=0.1419(1)$
\cite{ukqcd:strange}), and two values of the heavy quark mass
($\kappa_h=0.121,0.129$).  We employ two values of the $B$ meson
momentum ($(12a/\pi) {\vec{p}} =(0,0,0)$, $(1,0,0)$), and seventeen
values of the momentum, $\vec{q}$, injected at the operator, with
magnitudes between 0 and $2\pi/(12a)$. To improve statistics we
average over all equivalent momenta, and utilise the discrete symmetries $C$
and $P$, where possible.

Provided the three points in the correlators of
Eq.(~\ref{eq:3pt-correlator}) are sufficiently separated in time,
the ground state contribution to the ratio dominates:
\begin{equation}
\label{eq:3pt-asymptotic}
C_{\rho\mu\nu}  \mathop{\longrightarrow}\limits_{t,t_f - t\to\infty} 
 \frac{1}{Z_{P}Z_{V}} \sum_\epsilon
\epsilon_{\rho} 
\langle V(k,\epsilon) | \overline{s} \sigma_{\mu\nu} b
 | P(p) \rangle + \dots 
\end{equation} 
and $C_{\rho\mu\nu}$ approaches a plateau.
The factors $Z_{P}$, $Z_V$ and the energies of
the pseudoscalar and vector particles are obtained from fits to
two-point Euclidean correlators.

The form factor $T_1$ can be conveniently extracted from the matrix
elements by considering different components of the relation,
\begin{equation} 
4( k^\alpha p^\beta -  p^\alpha k^\beta) T_1(q^2) =
\varepsilon^{\alpha\beta\rho\mu} 
 \sum_\epsilon \epsilon_{\rho} 
 \langle V(k,\epsilon) | \overline{s} \sigma_{\mu\nu} b
 | P(p) \rangle q^\nu .
\end{equation}
We see a plateau in $T_1$ about $t=12$, and fit
$T_1(t;\vec{p},\vec{q})$ to a constant for $t=11,12,13$, where
correlations are maintained between all of the time slices.  The use
of smeared operators for the heavy quarks provides a very clean
signal, with stable plateaus forming before timeslice 11. Data with
initial or final momenta greater than $(\pi/12a)\sqrt{2}$ are
excluded, because they have larger statistical and systematic
uncertainties.

\begin{figure}
\pspicture{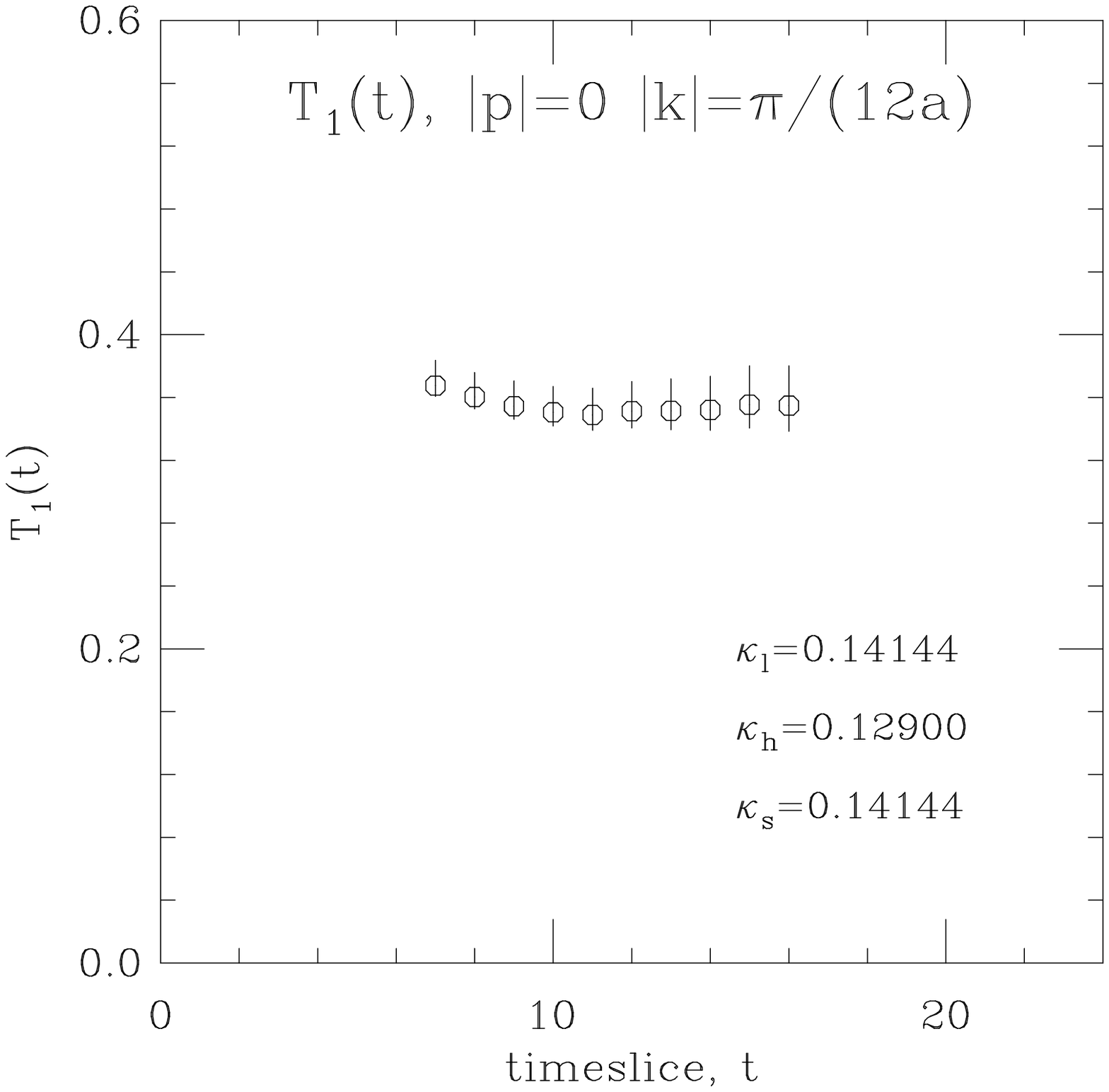}
\caption{$T_1$ vs. timeslice, $t$.
(For computational reasons, only timeslices 7---16 were stored) }
\label{fig:t1-vs-time} 
\end{figure}

The data for the heaviest of our light quarks,
$\kappa_l=\kappa_s=0.14144$, with the smallest statistical
errors, are shown in Fig.~\ref{fig:t1-vs-time}.

\section{Results}
\begin{figure}
\pspicture{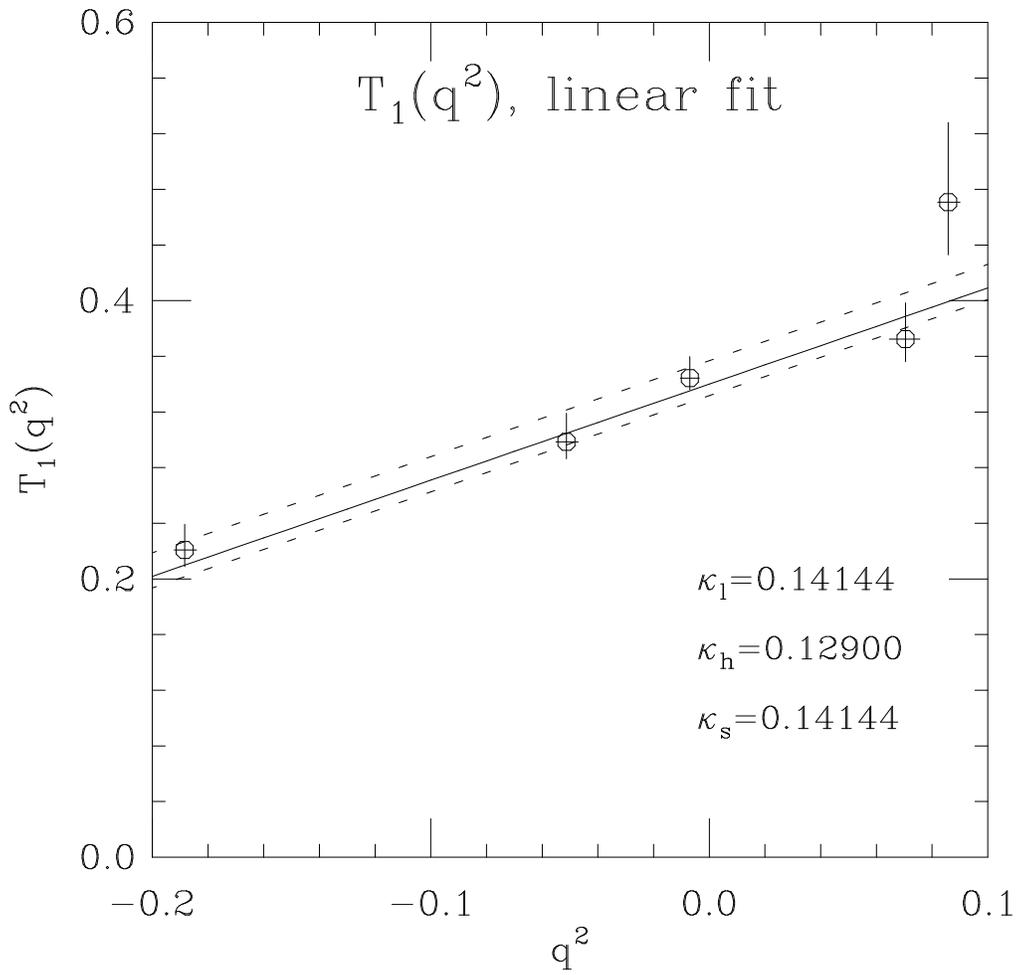}
\caption{$T_1(q^2)$, with a linear fit. The dotted lines represent
the 68\% confidence levels of the fit at $q^2=0$} 
\label{fig:t1-vs-q2}
\end{figure}

We fit $T_1(q^2)$ to a linear model in order to obtain the
on-shell form factor $T_1(q^2{=}0)$,
\begin{equation}
T_1(q^2)= a+b q^2,
\end{equation}
allowing for correlations between 
the energies of the vector and 
pseudoscalar particles and $T_1$ at each $q^2$.
For our range of masses and momenta the differences between
linear and pole model fits are small.  The data and fit for
$\kappa_l=\kappa_s=0.14144$ are shown in
Fig.~\ref{fig:t1-vs-q2}.

The light quark mass is set to zero by a correlated chiral
extrapolation to $\kappa_l=\kappa_{crit}(=0.14315(2))$ 
We assume that the
on-shell $T_1$ varies with the light kappa values, $\kappa_l$,
according to a linear model,
\begin{equation}  
T_1(\kappa_s,\kappa_h,\kappa_l)= T^{crit}_1(\kappa_s,\kappa_h)
+
\Delta_l(\kappa_s,\kappa_h)
\left(\frac{1}{\kappa_l}-\frac{1}{\kappa_{crit}}
\right).
\end{equation}
The
strange quark mass is set to its physical value by interpolation
$(\kappa_s=\kappa^{phys}_{s}=0.14183(5))$. 

Finally we perform an extrapolation from the two
pseudoscalar masses up to $m_B$ using a model motivated by heavy
quark effective theory,
\begin{equation}
T_1(q^2{=}0;m_{P})= A + \frac{B}{m_{P}} .
\end{equation} 

After performing this extrapolation, we obtain
$T_1(q^2{=}0;m_B)=\ukqcdresult$,
where the quoted error is purely statistical.
The
finite renormalization needed for the lattice--continuum matching of
the $\sigma_{\mu\nu}$ operator has been calculated
\cite{borrelli:improved-operators} but has a negligible effect here
($O(2\%)$) and is not included.

\begin{figure}
\pspicture{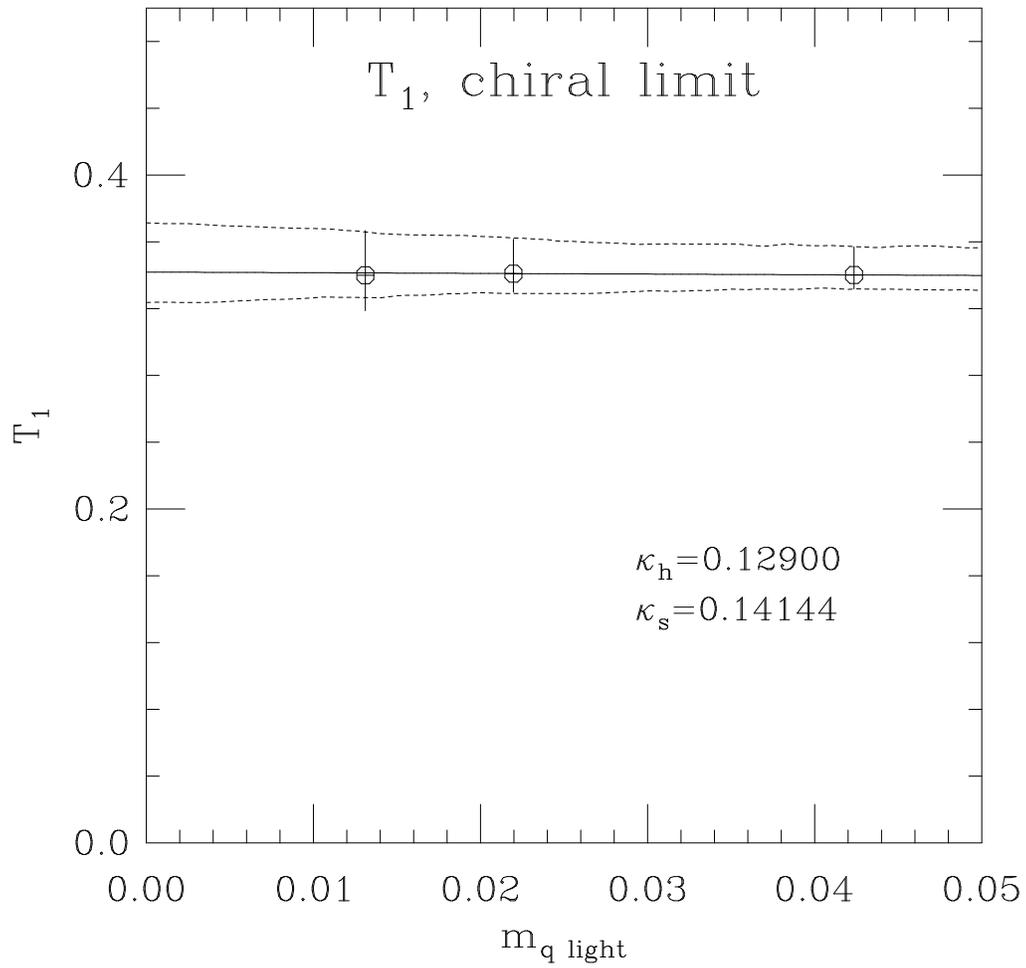}
\caption{Chiral extrapolation of $T_1(q^2{=}0)$.
The dotted lines indicate the 68\% confidence levels of the fit.
$m_{q~light}$ is the lattice pole mass. } 
\label{fig:t1-chiral-extrapolation} 
\end{figure}

We note that the slopes of the form factor $T_1$ with respect to
$\kappa_l$ in the chiral extrapolations are consistent with zero
(Fig.~\ref{fig:t1-chiral-extrapolation}), which indicates that
$T_1(\kappa_s,\kappa_h,\kappa_l)$ is almost independent of
$\kappa_l$. 
However, this behaviour occurs only for the 
spectator quark, and is not seen to the same extent for the
interacting strange quark.
We explore this 
by fitting $T_1$ to a constant for the three
values of $\kappa_l$. 
We find that the $\chi^2$ per degree of freedom 
is comparable to the original linear model,  
indicating that the model is statistically valid.
Using this approach,
the final statistical error is
significantly reduced, and we obtain
$T_1(q^2{=}0;m_B)=\ukqcdflatresult$.
The results for $T_1$, using both analysis procedures,
are shown in Fig.~\ref{fig:t1-at-mb}.

\section{Conclusions}

\begin{figure}
\pspicture{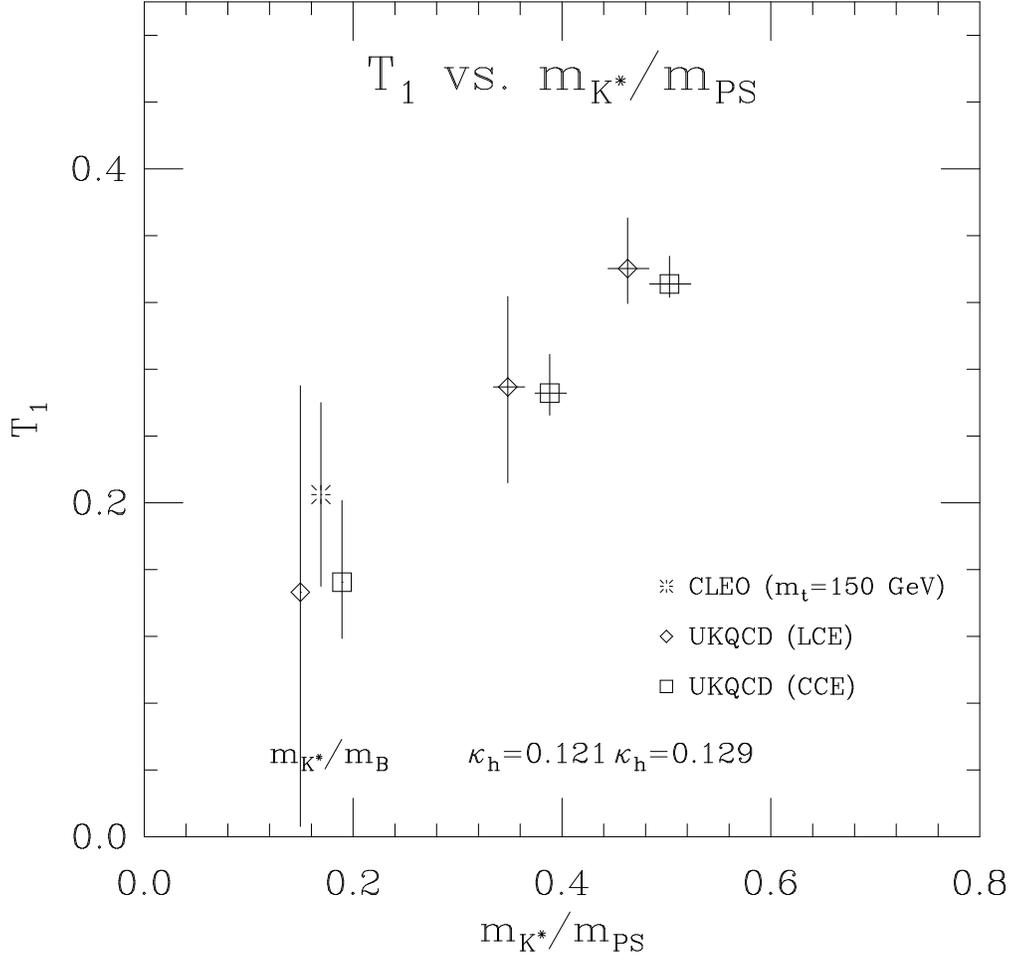}
\caption{Extrapolation of $T_1(q^2{=}0)$ to $m_B$.
LCE --- using linear chiral extrapolation, CCE --- using constant
chiral extrapolation for the spectator quark.  (N.B. for clarity, the
LCE and CCE points have been displaced horizontally by 0.02 to the
left and right respectively)}
\label{fig:t1-at-mb} 
\end{figure}

In this letter we have reported on an {\it ab initio} 
computation of the form factor for the decay $B \to K^* \gamma$.
The large number of gauge configurations used in this
calculation enables an extrapolation to the appropriate masses
to be made and gives a statistically meaningful result.  
To compare this
result with experiment we convert the preliminary branching ratio from
CLEO, $BR(\BtoKstargamma) = (4.5 \pm 1.5 \pm 0.9) \times 10^{-5}$
based on 13 events, into its corresponding $T_1$ form factor, assuming
the Standard Model.  We work at the scale $\mu=m_b=4.39\,\mbox{GeV}$
and use values from the Particle Data Book
\cite{pdb} combined with Eq.(~\ref{eq:decay-rate}).  Setting the mass
of the top quark to be $m_t=100$, $150$ and $200 \,\mbox{GeV}$ we find
$T^{\mbox{\it\scriptsize exp}}_1$ to be $\cleoone$, $\cleoresult$ and
$\cleotwo$ respectively.  The two lattice results are consistent with
these experimental numbers within statistical errors.  This is also
shown in Fig.~\ref{fig:t1-at-mb}.

%
%Systematics etc.
%
Although  systematic errors of this calculation resulting
from the quenched approximation, finite volume and other lattice
artefacts remain to be explored, 
we believe that we have shown the phenomenological utility
of the lattice for probing the limits of the Standard Model.

\vskip 0.5cm
\noindent 
{\bf Acknowledgments:} {
The authors wish to thank A.~Soni, C.~Bernard, A.~El-Khadra and 
members of the UKQCD collaboration, including C.~Allton for useful 
discussions on this topic. JMF thanks the Nuffield Foundation for 
support under the scheme of Awards for Newly Appointed Science 
Lecturers. The Wingate Foundation is acknowledged for its support of 
HPS by a scholarship.  DGR (Advanced Fellow) and DSH (Personal Fellow) 
acknowledge the support of the Science and Engineering Research 
Council.}

\clearpage 

\end{document}